\documentclass[12pt]{article}
\usepackage{amsmath,amssymb,theorem,cite,epsfig,url,psfrag,eepic,mathtools}
\advance \topmargin by -\headheight
\advance \topmargin by -\headsep     
\evensidemargin \oddsidemargin
\def\Maketitle{{\def\newpage{}\maketitle}}
\makeatletter
\def\Appendix{\appendix
  \def\@seccntformat##1{Appendix~\csname the##1\endcsname.~~}}
\makeatother
\makeatletter
\@addtoreset{equation}{section}

\makeatother

\def\Xint#1{\mathchoice
{\XXint\displaystyle\textstyle{#1}}%
{\XXint\textstyle\scriptstyle{#1}}%
{\XXint\scriptstyle\scriptscriptstyle{#1}}%
{\XXint\scriptscriptstyle\scriptscriptstyle{#1}}%
\!\int}
\def\XXint#1#2#3{{\setbox0=\hbox{$#1{#2#3}{\int}$}
\vcenter{\hbox{$#2#3$}}\kern-.5\wd0}}

\def\dashint{\Xint-}
\begin{document}
\rightline{RUNHETC-2013-13}
\title{\textbf{On spectrum of ILW hierarchy\\ in conformal field theory}\vspace*{.3cm}}
\date{}
\author{A.~V.~Litvinov$^{1,2}$\\[\medskipamount]
$^1$~\parbox[t]{0.85\textwidth}{\normalsize\it\raggedright
Landau Institute for Theoretical Physics,
142432 Chernogolovka, Russia}\vspace*{2pt}\\[\medskipamount]
$^2$~\parbox[t]{0.85\textwidth}{\normalsize\it\raggedright
NHETC, Department of Physics and Astronomy, Rutgers University,\\ Piscataway, NJ 08855-0849, USA}}
\Maketitle
\abstract{We consider a system of Integrals of Motion in conformal field theory related to the $\mathfrak{gl}(2)$ Intermediate Long Wave equation. It interpolates between the system studied by Bazhanov, Lukyanov and Zamolodchikov and  the one studied by the author and collaborators. We find Bethe anzatz equations for the spectrum of this system and its $\mathfrak{gl}(n)$ generalizations.}
\section{Introduction}
The importance of local Integrals of Motion (IM) in conformal field theory was emphasized  in \cite{Zamolodchikov:1989zs}. They are responsible for integrable perturbations of the theory. The major example -- the system of IM's which survives $\phi_{1,3}$ perturbation (quantum KdV system) was studied  in  \cite{Bazhanov:1994ft,Bazhanov:1996dr,Bazhanov:1998dq}. Later it was realized that  eigenvalues of Baxter's functions derived in \cite{Bazhanov:1994ft,Bazhanov:1996dr,Bazhanov:1998dq}  coincide with monodromy coefficients of certain second order differential operator \cite{Dorey:1998pt,Bazhanov:1998wj,Bazhanov:2004fk}. This phenomenon, known as ODE/IM correspondence, applies to variety of examples (for review see \cite{Dorey:2007zx}) and seems to be quite general. Recently, ODE/IM correspondence was pushed forward and generalized for the vacuum sector of certain massive integrable QFT's \cite{Lukyanov:2010rn,Dorey:2012bx,Lukyanov:2013wra}. 

New type of integrable systems in conformal field theory was studied in the paper \cite{Alba:2010qc} devoted to proof of AGT conjecture \cite{Alday:2009aq}. By analogy to quantum KdV  it can be called quantum  $\mathfrak{gl}(2)$ Benjamin-Ono system. It has very simple spectrum and its eigenfunctions  constitute a distinguished basis of states with remarkable property of factorization of matrix elements. Quantum KdV and $\mathfrak{gl}(2)$ BO integrable systems are both special limits of more general  one called quantum $\mathfrak{gl}(2)$ Intermediate Long Wave system, which is the main actor of this paper. Our original motivation was to extend ODE/IM correspondence for this case. For the time being this is still an open problem. Instead we found finite Bethe anzatz equations for the spectrum  without referring to any differential equation. Our BA equations are valid in the limiting KdV case and provide different description of the spectrum  compared to  \cite{Bazhanov:2004fk}.

This paper is organized as follows.  In section \ref{ILW-CFT} we define quantum $\mathfrak{gl}(2)$ ILW integrable system and describe equations for  its spectrum in two limiting cases: KdV and $\mathfrak{gl}(2)$ BO. In section \ref{spectrum} we formulate our main results: Bethe anzatz equations for the spectrum and determinant representation for norms of eigenfunctions. In section \ref{remarks} we give some remarks. In particular, we propose generalization of our BA equations for $W_{n}$ algebras. In appendix we describe semiclassical limit of  $\mathfrak{gl}(2)$ ILW system.
\section{ILW Integrals of Motion in CFT}\label{ILW-CFT}
We consider conformal field theory on a cylinder with the  symmetry algebra $\textrm{Vir}\oplus\textrm{H}$ (Virasoro and Heisenberg algebras) generated by the currents $T(x)$ and $J(x)$
\begin{equation}
   T(x)=-\frac{c}{24}+\sum_{n=-\infty}^{\infty}L_{n}e^{-inx},\qquad J(x)=\sum_{n\neq0}a_{n}e^{-inx}.
\end{equation}
The modes $L_{n}$ and $a_{n}$ satisfy commutation relations\footnote{We set zero mode of the field $J(x)$ to zero because it plays no essential role in  our construction.}
\begin{equation}\label{comm-relat}
   [L_{m},L_{n}]=(m-n)L_{m+n}+\frac{c}{12}(m^{3}-m)\delta_{m,-n},\quad
   [a_{m},a_{n}]=\frac{m}{2}\delta_{m,-n},\quad[L_{m},a_{n}]=0,
\end{equation}
where the central charge is parameterized in a standard way 
\begin{equation*}
   c=1+6Q^{2},\qquad Q=b+\frac{1}{b}.
\end{equation*}
In the universal enveloping algebra of $\textrm{Vir}\oplus\textrm{H}$ we are looking for the family of quantum Integrals of Motion $\mathbf{I}_{k}$ ($k=1,2,\dots$)
which commute among themselves $[\mathbf{I}_{k},\mathbf{I}_{l}]=0$ and start with
\begin{equation}\label{first-Quantum-densities}
  \begin{aligned}
     &\mathbf{I}_{1}=L_{0}+2\sum_{k=1}^{\infty}a_{-k}a_{k}-\frac{c+1}{24},\\
     &\mathbf{I}_{2}=\sum_{k\neq0}L_{-k}a_{k}+2iQ\sum_{k=1}^{\infty}k\coth(k\tau ) a_{-k}a_{k}+\frac{1}{3}\sum_{i+j+k=0}a_{i}a_{j}a_{k},
  \end{aligned}
\end{equation}
here $\tau$ is arbitrary complex parameter.
We note that the integral $\mathbf{I}_{2}$ is non-local due to the presence of operator $k\coth(k\tau)$. The integrals $\mathbf{I}_{k}$ for $k>2$ are all determined from the condition of their commutativity with $\mathbf{I}_{1}$ and $\mathbf{I}_{2}$. It is convenient to introduce densities $\mathbf{G}_{k}(x)$ by
 \begin{equation}\label{Quantum-IMs}
   \mathbf{I}_{k}=\frac{1}{2\pi}\int_{0}^{2\pi}\mathbf{G}_{k+1}(x)dx.
\end{equation}
The first four of these are
\begin{equation}\label{Quantum-densities}
  \begin{aligned}
      &\mathbf{G}_{2}=T+J^{2},\\
     &\mathbf{G}_{3}=TJ+iQ J\mathcal{D}J+\frac{1}{3}J^{3},\\
     &\mathbf{G}_{4}=T^{2}+6TJ^{2}+6iQT\mathcal{D}J+6iQ J^{2}\mathcal{D}J-6Q^{2}(\mathcal{D}J)^{2}+(1+Q^{2})J_{x}^{2}+J^{4},\\
     &\hspace*{-2pt}\begin{multlined}
      \mathbf{G}_{5}=T^{2}J+\frac{iQ}{2}T\mathcal{D}T-2Q^{2}T\mathcal{D}^{2}J+\frac{1}{2}T_{x}J_{x}+
      4iQTJ\mathcal{D}J+iQJ^{2}\mathcal{D}T+2TJ^{3}-2iQ^{3}\mathcal{D}J\mathcal{D}^{2}J+\\
      +iQ\left(\frac{Q^{2}}{2}-\frac{5}{6}\right)J_{xx}\mathcal{D}J-2Q^{2}J^{2}\mathcal{D}^{2}J+(1+Q^{2})JJ_{x}^{2}
      -2Q^{2}J(\mathcal{D}J)^{2}+\frac{4iQ}{3}J^{3}\mathcal{D}J+\frac{iQ}{2}J^{2}\mathcal{D}J^{2}+\frac{1}{5}J^{5}
     \end{multlined}
  \end{aligned}
\end{equation}
where $\mathcal{D}$ is the operator whose Fourier transform is $k\coth(k\tau)$ for $k\neq 0$ and $0$ for $k=0$.

The composite fields in the expressions \eqref{Quantum-densities}  require proper regularization. For the local fields (like $T^{2}$, $J^{4}$ etc) the regularization procedure is standard (see \cite{Bazhanov:1994ft} for details).  We call it analytic ordering. For the  monomial densities involving operator $\mathcal{D}$ we were unable to find any canonical way to define regularization. In \eqref{Quantum-densities} all non-local monomial densities of the form $(\mathcal{D}^{n}A)(\mathcal{D}^{m}B)\dots$ (where $A,B,\dots$ are local densities) are assumed to be Wick ordered. This prescription (analytic ordering for local densities and Wick ordering  for non-local ones) does not work for higher densities however. Starting from some level one has to add counterterms in order to ensure commutativity with $\mathbf{I}_{2}$.  The structure of these counterterms is not clear at the moment. We have explicitly constructed them up to spin $k=7$ which seems to be quite good confirmation of the existence of an infinite tower of IM's. Moreover, in the limit $b\rightarrow0$ the system \eqref{first-Quantum-densities}--\eqref{Quantum-densities} reduces to the classical $\mathfrak{gl}(2)$  Intermediate Long Wave system which is  known to be integrable \cite{Lebedev-Radul} (see more details in appendix \ref{D}).

In this paper we study the spectrum of the integrable system \eqref{first-Quantum-densities}--\eqref{Quantum-densities}. Consider highest weight representation $\pi_{P}$ of  $\textrm{Vir}\oplus\textrm{H}$ with the highest weight $|P\rangle$ defined by
\begin{equation}
   L_{n}|P\rangle=a_{n}|P\rangle=0,\quad\text{for}\quad n>0,\qquad L_{0}|P\rangle=\Delta|P\rangle,\quad\text{with}\quad \Delta=\frac{Q^{2}}{4}-P^{2}.
\end{equation}
Then the basis in $\pi_{P}$ is formed by monomials
\begin{equation}\label{naive-basis}
   a_{-k_{1}}\dots a_{-k_{m}}L_{-l_{1}}\dots L_{-l_{m}}|P\rangle,
\end{equation}
with $k_{1}\geq k_{2}\geq\dots\geq k_{m}$ and $l_{1}\geq l_{2}\geq\dots\geq l_{n}$. The eigenspaces of the operator $\mathbf{I}_{1}$ are spanned by the vectors \eqref{naive-basis} with
\begin{equation*}
   k_{1}+\dots+k_{m}+l_{1}+\dots+l_{n}=N,
\end{equation*}
for any $N\geq0$. The number of states $\mathfrak{n}(N)$ at given level $N$  equals to the number of bipartitions of $N$ ($\mathfrak{n}(1)=2$, $\mathfrak{n}(2)=5$ etc.). The operators $\mathbf{I}_{k}$ with $k>1$ lift the degeneracy. We consider the spectral problem
\begin{equation}\label{spectral-problem}
    \mathbf{I}_{k}|P,N,j\rangle=\mathfrak{h}_{k}^{(N)}(\tau,P)|P,N,j\rangle,\quad j=1,\dots,\mathfrak{n}(N),
\end{equation}
where we have chosen the following normalization for the eigenvectors
\begin{equation}\label{Bethe-normalization}
 |P,N,j\rangle=(L_{-1}^{N}+\dots)|P\rangle.
\end{equation}
Missed terms in \eqref{Bethe-normalization} have a degree in $L_{-1}$ at most $N-1$.
The spectral problem \eqref{spectral-problem}  can be solved by linear algebra methods for any given value of $N$. But we are looking for some universal description of the spectrum valid for any $N$. Such a description is known in two limiting cases $\tau=0$ and $\tau=\pm\infty$. We call these limits KdV and BO limits respectively. Consider both limits in details.
\subsection{KdV limit}
In the limit $\tau\rightarrow0$ the integrable system  \eqref{first-Quantum-densities}--\eqref{Quantum-densities} separates into two. 
Namely, the operator $\mathcal{D}$ has the following expansion
\begin{equation}
   \mathcal{D}=\frac{1}{\tau}-\frac{\tau}{3}\partial^{2}-\frac{\tau^{3}}{45}\partial^{4}+O(\tau^{5}),
\end{equation}
and  $\mathbf{I}_{k}$'s behave as follows
\begin{equation}
\begin{aligned}
  &   \mathbf{I}_{2}=iQ\tau^{-1}\mathbf{I}_{1}^{\scriptscriptstyle{\textrm{free}}}+\dots,\qquad\qquad
  \mathbf{I}_{3}-6iQ\tau^{-1}\mathbf{I}_{2}=4iQ\tau^{-1}\mathbf{I}_{2}^{\scriptscriptstyle{\textrm{free}}}+\dots,\\
    &   \mathbf{I}_{4}-\frac{5iQ}{6}\tau^{-1}\mathbf{I}_{3}-3Q^{2}\tau^{-2}\mathbf{I}_{2}=
    iQ\tau^{-1}\left(\mathbf{I}_{3}^{\scriptscriptstyle{\textrm{free}}}-\frac{1}{3}\mathbf{I}_{3}^{\scriptscriptstyle{\textrm{BLZ}}}\right)+\dots,\\
\end{aligned}
\end{equation}
where $\mathbf{I}_{2n-1}^{\scriptscriptstyle{\textrm{BLZ}}}$ are the Integrals of Motion considered by Bazhanov, Lukyanov and Zamolodchikov \cite{Bazhanov:1998dq,Bazhanov:1996dr,Bazhanov:1994ft}
\begin{equation}\label{BLZ-system}
      \mathbf{I}_{1}^{\scriptscriptstyle{\textrm{BLZ}}}=L_{0}-\frac{c}{24},\qquad
      \mathbf{I}_{3}^{\scriptscriptstyle{\textrm{BLZ}}}=2\sum_{k>0}L_{-k}L_{k}+L_{0}^{2}-\frac{c+2}{12}L_{0}+\frac{c(5c+22)}{2880},\dots\dots 
\end{equation}
and $\mathbf{I}_{n}^{\scriptscriptstyle{\textrm{free}}}$ are the Integrals of Motion for the free field
\begin{equation}\label{Free-system}
    \mathbf{I}_{1}^{\scriptscriptstyle{\textrm{free}}}=2\sum_{k=1}^{\infty}a_{-k}a_{k},\quad \mathbf{I}_{2}^{\scriptscriptstyle{\textrm{free}}}=\frac{1}{3}\sum_{i+j+k=0}a_{i}a_{j}a_{k},\dots\dots
\end{equation}
One can show that the system \eqref{Free-system} is equivalent to BLZ system \eqref{BLZ-system} at the ``free fermion'' point $c=-2$. 

Below we consider only the system \eqref{BLZ-system}. The spectrum of \eqref{BLZ-system} on higher level states was found in \cite{Bazhanov:2004fk}. In view of ODE/IM correspondence, it is described in terms of second order differential operator
\begin{equation}\label{BLZ-oper}
   (-\partial^{2}+t(z)+\lambda z^{\varkappa})\psi(z)=0,
\end{equation}
where
\begin{equation}
   t(z)=\frac{l(l+1)}{z^{2}}+\frac{1}{z}\left(1-\sum_{k=1}^{N}\gamma_{k}\right)+\sum_{k=1}^{N}\left(\frac{2}{(z-z_{k})^{2}}+\frac{\gamma_{k}}{z-z_{k}}\right),
\end{equation}
with $l=-\frac{1}{2}-bP$, $\varkappa=-2-b^{2}$ and $N$ coincides with the level (eigenvalue of $L_{0}-\Delta$). One demands that the operator \eqref{BLZ-oper} has no monodromy at points $z_{k}$ for all values of the spectral parameter $\lambda$. This condition is equivalent to the system of equations
\begin{equation}\label{monodromy-free-equations}
   \frac{c_{1,k}^{3}}{4}-c_{1,k}c_{2,k}+c_{3,k}=0,
\end{equation}
where $c_{j,k}$ are the coefficients of the expansion
\begin{equation*}
   t(z)+\lambda z^{\varkappa}=\frac{2}{(z-z_{k})^{2}}+\frac{c_{1,k}}{z-z_{k}}+c_{2,k}+c_{3,k}(z-z_{k})+O((z-z_{k})^{2})\quad\text{at}\quad
   z\rightarrow z_{k}.
\end{equation*}
The equations  \eqref{monodromy-free-equations} are linear in $\lambda$. At degree $1$ one has $\gamma_{k}=\frac{\varkappa}{z_{k}}$,
while at degree zero we get a system of  equations (here $\Delta=\frac{Q^{2}}{4}-P^{2}$)
\begin{equation}\label{BA-BLZ}
   \frac{b^{4}\Delta}{z_{j}}+\sum_{k\neq j}R(z_{j},z_{k})=1+b^{2},\qquad j=1,\dots, N,
\end{equation}
with
\begin{equation*}
    R(x,y)=\frac{b^4x^2+\left(2+b^2\right)\left(1-2b^2\right)xy+\left(1+b^2\right) \left(2+b^2\right) y^2}{(x-y)^{3}}.
\end{equation*}
It was conjectured in  \cite{Bazhanov:2004fk} that eigenvalues of the operators  $\mathbf{I}_{2n-1}^{\scriptscriptstyle{\textrm{BLZ}}}$ are symmetric polynomials of $z_{k}$'s. In particular,  for eigenvalues of $\mathbf{I}_{3}^{\scriptscriptstyle{\textrm{BLZ}}}$ and $\mathbf{I}_{5}^{\scriptscriptstyle{\textrm{BLZ}}}$ one has
\begin{equation}\label{I3-I5-BLZ}
\begin{aligned}
  &  \mathcal{I}_{3}=\mathcal{I}_{3}^{\scriptscriptstyle{\textrm{vac}}}(\Delta+N)+\frac{4(1+b^{2})}{b^{4}}\sum_{k=1}^{N}z_{k},\\
   &\mathcal{I}_{5}=\mathcal{I}_{5}^{\scriptscriptstyle{\textrm{vac}}}(\Delta+N)+
   \frac{(2-b^2)(1+b^2)}{b^4}\left(\frac{12(\Delta+N-1)}{2-3b^2}-\frac{(1-2 b^2)}{2 b^2}\right)
   \sum_{k=1}^{N}z_{k}
   -\frac{24 \left(1+b^2\right)^2}{b^6 \left(2-3 b^2\right)}\sum_{k=1}^{N}z_{k}^{2},
\end{aligned}
\end{equation}
where $\mathcal{I}_{2n-1}^{\scriptscriptstyle{\textrm{vac}}}(\Delta+N)$ are the vacuum eigenvalues. We note that there is a formal similarity between  \eqref{BA-BLZ} and Bethe anzatz equations for the Gaudin model (see more in  \cite{Feigin:2007fk}). Unfortunately, nothing is known about generalizations of eqs. \eqref{BA-BLZ} for other models of CFT.
\subsection{BO limit}
In the limit $\tau\rightarrow\pm\infty$ the operator $\mathcal{D}$ degenerates to $\pm|k|$. The spectrum is very simple  \cite{Alba:2010qc} and coincides with the spectrum of two Calogero-Sutherland models. Consider the case $\tau\rightarrow\infty$. Take bipartition 
$(\lambda,\mu)=(\lambda_{1}\geq\lambda_{2}\geq\dots,\mu_{1}\geq\mu_{2}\geq\dots)$ with $|\lambda|+|\mu|=N$, then eigenvalues  corresponding to  $(\lambda,\mu)$ are linear combinations of 
\begin{equation}\label{h2-Calogero}
    h_{\lambda,\mu}^{(k)}(P)\overset{\text{def}}{=}h_{\lambda}^{(k)}(P)+ h_{\mu}^{(k)}(-P),
\end{equation}
where
\begin{equation}\label{h-Calogero}
   h_{\lambda}^{(k)}(P)=b^{1-k}\sum_{j=1}^{\infty}\left[\left(bP-\frac{b^{2}}{2}+\lambda_{j}+jb^{2}\right)^{k}-\left(bP-\frac{b^{2}}{2}+jb^{2}\right)^{k}\right].
\end{equation}
In particular, in the limit $\tau\rightarrow\infty$ eigenvalues of the operators $\mathbf{I}_{1}$, $\mathbf{I}_{2}$, $\mathbf{I}_{3}$ and $\mathbf{I}_{4}$ (defined in \eqref{first-Quantum-densities}--\eqref{Quantum-densities}) are given by
\begin{equation}
  \begin{aligned}
     &\mathbf{I}_{1}\sim h_{\lambda,\mu}^{(1)}(P)-\frac{c+1}{24},\\
     &\mathbf{I}_{2}\sim -\frac{i}{2}h_{\lambda,\mu}^{(2)}(P),
   \end{aligned}\qquad
   \begin{aligned}
     &\mathbf{I}_{3}\sim -2h_{\lambda,\mu}^{(3)}(P)-\frac{1+b^{2}}{2}N+\left(\Delta ^2-\frac{c+5}{12}\Delta +\frac{5 c^2+52c+15}{2880}\right),\\
     &\mathbf{I}_{4}\sim -\frac{i}{2}h_{\lambda,\mu}^{(4)}(P)-\frac{i}{4}(1+b^{2})h_{\lambda,\mu}^{(2)}(P),\\
  \end{aligned}
\end{equation}
The eigenvalues for the Calogero-Sutherland model usually appear in the literature in the form \eqref{h-Calogero}. We will use equivalent form which is more suitable for our purposes. Namely, for each box in the partition $\lambda$ with coordinates $(i,j)$ one associates a number (content of the box)
\begin{equation}\label{Bethe-root-Calogero} 
    x=P-\frac{Q}{2}+ib^{-1}+jb.
\end{equation}
Then the eigenvalues \eqref{h-Calogero} are symmetric polynomials in $x_{k}$
\begin{equation}\label{hnew-Calogero}
   h_{\lambda}^{(2)}(P)=2\sum_{k=1}^{|\lambda|}x_{k},\quad
   h_{\lambda}^{(3)}(P)=3\sum_{k=1}^{|\lambda|}x_{k}^{2}+\frac{|\lambda|}{4b^{2}},\quad
   h_{\lambda}^{(4)}(P)=4\sum_{k=1}^{|\lambda|}x_{k}^{3}+\frac{1}{b^{2}}\sum_{k=1}^{|\lambda|}x_{k},\quad\dots
\end{equation}
Contents for the partition $\mu$ are the same as \eqref{Bethe-root-Calogero} but with $P\rightarrow-P$ (see \eqref{h2-Calogero}).
So,  any bipartition $(\lambda,\mu)$ with $|\lambda|+|\mu|=N$ is characterized by $N$  roots of the form  \eqref{Bethe-root-Calogero} ($+P$ for partition $\lambda$ and $-P$ for $\mu$). Eigenvalues of IM's are linear combinations of power-sum symmetric polynomials of these roots (as follows from \eqref{h2-Calogero} and \eqref{hnew-Calogero}). 

Note that all sets of  roots described above solve the following system of equations
\begin{equation}\label{BA-BO}
   A(x_{i})\prod_{j\neq i}F(x_{i}-x_{j})=0\quad\text{for}\quad i=1,\dots,N,
\end{equation}
with
\begin{equation}\label{A-F}
   A(x)=\left(x+P-\frac{Q}{2}\right)\left(x-P-\frac{Q}{2}\right)\quad\text{and}\quad
   F(x)=\frac{(x-b)(x-b^{-1})}{x(x-Q)}.
\end{equation}
Indeed, it is obvious from \eqref{BA-BO} that at least one of the roots (say $x_{1}$) should be a solution to $A(x)=0$. For example, take $x_{1}=P+\frac{Q}{2}$.  Then the rest satisfy
\begin{equation*}
   A(x_{i})F(x_{i}-P-\frac{Q}{2})\prod_{j\neq 1,i}F(x_{i}-x_{j})=0\quad\text{for}\quad i=2,\dots,N.
\end{equation*}
Again, one of the roots (say $x_{2}$) should be a solution to $A(x)F(x-P-\frac{Q}{2})=0$. This equation has three roots $P+\frac{Q}{2}+b^{-1}$, $P+\frac{Q}{2}+b$ (compare to \eqref{Bethe-root-Calogero} ) and $-P+\frac{Q}{2}$. Proceeding exactly the same way one finds that all roots of \eqref{BA-BO} are of the form
\begin{equation*}
   P-\frac{Q}{2}+ib^{-1}+jb,\;(i,j)\in\lambda\quad\text{or}\quad
   -P-\frac{Q}{2}+ib^{-1}+jb,\;(i,j)\in\mu,
\end{equation*}
where $(\lambda,\mu)$ is some bipartition of $N$. The fact that $(\lambda,\mu)$ is actually a bipartition is controlled by two terms in denominator of $F(x)$. We note that the functions $A(x)$ and $F(x)$ are essential blocks for the contour integrals introduced in  \cite{Moore:1997dj,Lossev:1997bz}. The expansion of these integrals in terms of partitions is well known \cite{Nekrasov:2002qd}. 
\section{Spectrum}\label{spectrum}
For general values of the parameter $\tau$ we propose that the spectrum of integrable system \eqref{first-Quantum-densities}--\eqref{Quantum-densities} is governed by the system of Bethe anzatz equations\footnote{The same equations appeared  in \cite{NO-ILW}. We thank Mikhail Bershtein and Nikita Nekrasov for bringing this to our attention.}
\begin{equation}\label{BA-general}
   e^{\tau}A(x_{i})\prod_{j\neq i}F(x_{i}-x_{j})=e^{-\tau}A(x_{i}+Q)\prod_{j\neq i}F(x_{j}-x_{i})\quad\text{for}\quad i=1,\dots,N,
\end{equation} 
with $A(x)$ and $F(x)$ given by \eqref{A-F}\footnote{We note that in the limit $\tau\rightarrow\infty$ these equations are reduced to \eqref{BA-BO} (and similarly for $\tau\rightarrow-\infty$).}. Namely, the following holds:
\begin{enumerate}
\item The number of orbits of solutions to \eqref{BA-general} is exactly the number of bipartitions of $N$. We note that due to denominators $\prod_{i<j}1/x_{ij}$ in \eqref{BA-general} the Bethe roots are always pairwise distinct.
\item The eigenvalues $\mathfrak{h}_{k}^{(N)}(\tau,P)$ for the operators $\mathbf{I}_{k}$ \eqref{Quantum-IMs}--\eqref{Quantum-densities} on level $N$ are symmetric polynomials of Bethe roots
\begin{equation}
\begin{gathered}
  \mathfrak{h}_{2}^{(N)}(\tau,P)=-i\sum_{j=1}^{N}x_{j},\quad
  \mathfrak{h}_{3}^{(N)}(\tau,P)=-6\sum_{j=1}^{N}x_{j}^{2}+\frac{1-Q^{2}}{2}N+\left(P^4+\frac{P^2}{2}+\frac{Q^2+6}{240}\right),\\
  \mathfrak{h}_{4}^{(N)}(\tau,P)=2i\left(\sum_{j=1}^{N}x_{j}^{3}-\frac{1-Q^{2}}{4}\sum_{j=1}^{N}x_{j}\right),\quad
  \dots
\end{gathered}
\end{equation}
where $x_{i}$ are solutions to \eqref{BA-general}.
\item The norms of eigenvectors normalized as \eqref{Bethe-normalization} admit  standard determinant representation \cite{Korepin:1982fk,Slavnov:1989kx}
\begin{equation}\label{norm}
   \text{Norm}=\prod_{k=1}^{N}Q^{-1}A(x_{k})A(x_{k}+Q)\prod_{i\neq j}F(x_{i}-x_{j})
   \det\left(\frac{\partial \mathcal{Y}(x_{1},\dots,x_{N})}{\partial x_{i}\partial x_{j}}\right),
\end{equation}
where $\mathcal{Y}(x_{1},\dots,x_{N})$ is the Yang-Yang function \cite{Yang:1966ty}
\begin{multline}\label{YY}
  \mathcal{Y}(x_{1},\dots,x_{N})=\\
  =\sum_{k=1}^{N}\left[\omega\left(x_{k}+P-\frac{Q}{2}\right)+\omega\left(x_{k}-P-\frac{Q}{2}\right)-\omega\left(x_{k}+P+\frac{Q}{2}\right)
  -\omega\left(x_{k}-P+\frac{Q}{2}\right)\right]+\\+
  \sum_{i\neq j}\left[\omega(x_{i}-x_{j}-b)+\omega(x_{i}-x_{j}-b^{-1})+\omega(x_{i}-x_{j}+Q)\right]+
  2\tau\sum_{k=1}^{N}x_{k},
\end{multline}
with    $\omega(x)=x\left(\log x-1\right)$.
\end{enumerate} 
We have checked these statements by explicit computations up to level $4$. 

Sometimes it is convenient to rewrite Bethe anzatz equations \eqref{BA-general} in equivalent form which is  a  non-linear version of  celebrated Baxter's TQ equation \cite{Baxter-book}
\begin{equation}\label{TQ-equation}
   \mathbb{T}(x)\mathbb{Q}(x)=e^{\tau}A(x)\mathbb{Q}(x-b)\mathbb{Q}(x-b^{-1})\mathbb{Q}(x+Q)+
   e^{-\tau}A(x+Q)\mathbb{Q}(x+b)\mathbb{Q}(x+b^{-1})\mathbb{Q}(x-Q).
\end{equation}
Here $\mathbb{Q}(x)=\prod_{k=1}^{N}(x-x_{k})$ and $\mathbb{T}(x)$ is some polynomial. As far as we concerned Bethe anzatz equations similar to \eqref{BA-general} and \eqref{TQ-equation} (with rational $S$-matrix having multiple zeroes) have not been studied in the literature until recently \cite{2012arXiv1205.2968K}. In fact, in \cite{2012arXiv1205.2968K} they were studied without referring to a specific  integrable model.
\section{Concluding remarks}\label{remarks} 
In this paper we found Bethe anzatz equations for the spectrum of integrable system \eqref{first-Quantum-densities}--\eqref{Quantum-densities}. Our main results are presented in section \ref{spectrum}. Below we will make few remarks:
\begin{enumerate}
\item Bethe anzatz equations \eqref{BA-general} have smooth limit at $\tau\rightarrow 0$ where they describe the spectrum of BLZ system \eqref{BLZ-system}. One can show that  for any solution of  \eqref{BA-general} with $\tau=0$ one has
\begin{equation*}
   \sum_{k=1}^{N}x_{k}^{2p-1}=0\quad\text{for}\quad p\in \mathbb{Z},
\end{equation*}
so that eigenvalues for IM's with even spin vanish. For odd spin IM's one has  
\begin{equation}\label{I3-I5-BLZ2}
\begin{aligned}
  &\mathcal{I}_{3}=\mathcal{I}_{3}^{\scriptscriptstyle{\textrm{vac}}}(\Delta+N)+4N\Delta+2N(N-1)+12\sum_{k=1}^{N}x_{k}^{2},\\
  &
  \begin{multlined}
      \mathcal{I}_{5}=\mathcal{I}_{5}^{\scriptscriptstyle{\textrm{vac}}}(\Delta+N)+12N\Delta^{2}+\left(22N^{2}-\frac{(c+104)N}{6}\right)\Delta
      -\frac{N(N-1)}{12}(c-88N+76)+
      \\+
  \left(60(\Delta+N-1)+\frac{5(c-4)}{6}\right)\sum_{k=1}^{N}x_{k}^{2}+40\sum_{k=1}^{N}x_{k}^{4},
  \end{multlined}
\end{aligned}
\end{equation}
and similar (but more messy) expressions for higher spins. We note here very interesting phenomenon. Namely, the spectrum of BLZ system is described by two types of Bethe anzatz equations \eqref{BA-BLZ} and \eqref{BA-general} (with $\tau=0$). The relation between Bethe roots $z_{k}$ and $x_{k}$ can be deduced from \eqref{I3-I5-BLZ} and \eqref{I3-I5-BLZ2}. We think that this is similar to bispectral duality between Gaudin and XXX models \cite{Mukhin:2006fk}. Unfortunately, we were unable to find deformation of the equations \eqref{BA-BLZ} for $\tau\neq0$, but we believe that such a generalization exists. Recent results obtained in \cite{Mironov:2012ba,Bulycheva:2012ct, Gaiotto:2013bwa} could be useful here.
\item Bethe anzatz equations \eqref{BA-general} admit simple generalization for the case of $\textrm{W}_{n}$ algebras \cite{Zamolodchikov:1985wn}. For example, consider the algebra $\textrm{W}_{3}\oplus\textrm{H}$. Consider the system which starts with first non-trivial integral (here we use notations as in \cite{Fateev:2007ab})
\begin{equation}\label{I2-W3}
\mathbf{I}_{2}=\sqrt{\frac{2}{3}}\left(\sum_{k\neq0}L_{-k}a_{k}+6iQ\sqrt{\frac{3}{8}}\sum_{k>0}k\coth(k\tau ) a_{-k}a_{k} +
\frac{\sqrt{4+15Q^{2}}}{4}W_{0}+\frac{1}{3}\sum_{i+j+k=0}a_{i}a_{j}a_{k}\right). 
\end{equation}
One can argue that \eqref{I2-W3} generates an infinite system of commuting Integrals of Motion. 
In the limit $\tau\rightarrow0$ this system reduces to quantum Boussinesq system considered in \cite{Bazhanov:2002uq}, while for $\tau=\pm\infty$ one arrives to one studied in \cite{Fateev:2011hq}. Based on explicit calculations on lower levels, we propose that the spectrum is described by Bethe anzatz equations
\begin{equation}\label{BA-general-3}
   e^{\tau}A_{3}(x_{i})\prod_{j\neq i}F(x_{i}-x_{j})=e^{-\tau}A_{3}(x_{i}+Q)\prod_{j\neq i}F(x_{j}-x_{i})\quad\text{for}\quad i=1,\dots,N,
\end{equation} 
where $F(x)$ is the same as in \eqref{A-F},
\begin{equation}
   A_{3}(x)=\left(x+P_{1}-\frac{Q}{2}\right)\left(x+P_{2}-\frac{Q}{2}\right)\left(x+P_{3}-\frac{Q}{2}\right),
\end{equation} 
and $P_{k}=(P,h_{k})$ are projection of the momenta on weights of first fundamental representation of $\mathfrak{sl}(3)$ (note that $P_{1}+P_{2}+P_{3}=0$). In particular, eigenvalues of \eqref{I2-W3} are given by
\begin{equation}
   \mathbf{I}_{2}\sim\mathcal{I}_{2}^{\textrm{vac}}+i\sum_{k=1}^{N}x_{k},
\end{equation}
where $\mathcal{I}_{2}^{\textrm{vac}}$ is the vacuum eigenvalue. Expressions for norms of Bethe states are similar to \eqref{norm}.
Similarly to the case of $n=2$ one recovers quantum Boussinesq system \cite{Bazhanov:2002uq} in the limit $\tau\rightarrow0$. In particular, eigenvalues of the operator $W_{0}$ are given by $w_{0}-2\sum_{k}x_{k}$, where $x_{k}$ satisfy \eqref{BA-general-3} with $\tau=0$. We note that the ``Gaudin'' type BA equations for the quantum Boussinesq system (similar to \eqref{BA-BLZ}) presently are not known and hence our BA equations is the only description of the spectrum of this integrable model. 

Generalization of equations \eqref{BA-general} and \eqref{BA-general-3} for arbitrary rank $n$ is straightforward. Only the function $A(x)$ changes to 
\begin{equation}
   A_{n}(x)=\prod_{k=1}^{n}\left(x+P_{k}-\frac{Q}{2}\right).
\end{equation}
We propose that corresponding BA equations describe the spectrum of $\mathfrak{gl}(n)$ generalization of ILW integrable system built on the symmetry algebra $\textrm{W}_{n}\oplus\textrm{H}$. This system is generated by spin $2$ integral $\mathbf{I}_{2}$ similar to  \eqref{I2-W3}. Its explicit form in the limit $\tau\rightarrow\infty$ can be found in \cite{Estienne:2011qk,Maulik:2012wi}.
\item We believe that there exists determinant representation for matrix elements of local operators between Bethe states (formfactors) similar to  \eqref{norm}. Such a representation is a peculiar property of systems described by Bethe anzatz equations \cite{Korepin:1982fk,Slavnov:1989kx}. 
\item It is a challenging problem to find how our problem fits in general framework of quantum inverse scattering method. In particular,  it is tempting to construct $\mathbb{T}$ and $\mathbb{Q}$ operators whose eigenvalues satisfy \eqref{TQ-equation}. We note that Baxter's functions $\mathbb{T}(x)$ and $\mathbb{Q}(x)$ are qualitatively different from whose defined and studied in \cite{Bazhanov:1994ft,Bazhanov:1996dr,Bazhanov:1998dq}. The latter are entire functions with infinite number of zeroes, while $\mathbb{T}(x)$ and $\mathbb{Q}(x)$ are polynomials whose degree is determined by the level $N$.
\item We find it curious that the spectrum of CFT, which is a continuous theory, is governed by finite Bethe anzatz equations.
This is to be compared to the TBA equations from \cite{Bazhanov:1994ft,Bazhanov:1996dr,Bazhanov:1998dq} which are obtained from a discrete spin chain in a scaling limit (see  \cite{Boos:2009fs} for the recent progress in this direction). We think that this phenomenon, the appearance of finite spin chain equations in a continuous theory, is very similar  to one  observed in AdS/CFT integrability \cite{Beisert:2010jr}. 
\item Unpleasant property  of  ILW system considered in this paper is that it corresponds to integrable perturbation of CFT by some non-local operator. 
At the moment it is not clear whether such a perturbation defines self-consistent QFT or not.  We note that  the non-locality parameter $\tau$ in BA equations \eqref{BA-general} plays the same role as a twist parameter in spin chains. From this point of view (very naive in fact) introduction of $\tau$ looks natural. We hope that corresponding massive integrable QFT, if exists,  could have interesting physical applications. 
\item We think that many interesting models of CFT admit description of their spectra in terms of Bethe anzatz equations similar to \eqref{BA-general}. One of the immediate candidates would be theory with para-Virasoro symmetry \cite{Bershtein:2010wz} and   more general models studied recently  in \cite{Belavin:2013fk}. We expect that spectra of corresponding integrable systems are described by some variants of nested Bethe anzatz equations. 
\end{enumerate}
We will address questions formulated above in future publications.
\section*{Acknowledgements}
The author thanks Vladimir Bazhanov, Vladimir Fateev, Sergei Lukyanov, Feodor Smirnov and Alexander Zamolodchikov for  explanations, discussions of the subject and their interest to this work. He also thanks Grisha Tarnopolsky for collaboration at the initial stage of this project.  

This work was supported by DOE grant \# DE-FG02-96 ER 40949, by RFBR grants 12-02-01092, 12-02-33011, 13-01-90614 and by Russian Ministry of Education and Science grants 2012-1.5-12-000-1011-012  and 2012-1.1-12-000-1011-016.
\Appendix
\section{Semiclassical limit of $\mathfrak{gl}(2)$ ILW system}\label{D}
Consider the limit $b\rightarrow 0$ with
\begin{equation*}
   T(x)\rightarrow-\frac{1}{b^{2}}u(x),\qquad J(x)\rightarrow-\frac{i}{b}v(x),\qquad
   [,]\rightarrow-2\pi b^{2}\{,\}
\end{equation*}
then the quantum IM's \eqref{Quantum-IMs} tend to classical IM's (conserved quantities) for $\mathfrak{gl}(2)$ Intermediate Long Wave equation \cite{Lebedev-Radul}
\begin{equation}\label{ILW-equation}
    \begin{cases}
     u_{t}+vu_{x}+2uv_{x}+\frac{1}{2}v_{xxx}=0,\\
     v_{t}+\frac{u_{x}}{2}+\mathcal{T}v_{xx}+vv_{x}=0,
   \end{cases}
\end{equation}
where  $\mathcal{T}$ is non-local operator given  by the principal value integral\footnote{Operator $\mathcal{D}$ used in section \ref{ILW-CFT} is expressed as $\mathcal{D}=-\mathcal{T}\partial_{x}$.}
\begin{equation}
   \mathcal{T}F(x)\overset{\text{def}}{=}\frac{1}{2\pi}\,\dashint_{0}^{2\pi}F(y)\frac{\Theta_{1}'(\frac{y-x}{2}|q)}{\Theta_{1}(\frac{y-x}{2}|q)}\,dy,
\end{equation}
and $\Theta_{1}'(z|q)$ is the elliptic Theta function with $q=e^{-\tau}$.
The operator $\mathcal{T}$ has the following expansion at $\tau\rightarrow0$ and $\tau\rightarrow\infty$
\begin{equation}
\begin{aligned}
  &\mathcal{T}=-\frac{1}{\tau}\partial^{-1}+\frac{\tau}{3}\partial+\frac{\tau^{3}}{45}\partial^{3}+O(\tau^{5}),\quad \tau\rightarrow0,\\
  &\mathcal{T}=\mathcal{H}+O(\tau^{-2}),\quad\tau\rightarrow\infty,
\end{aligned}
\end{equation}
where $\mathcal{H}$ is the operator of Hilbert transform. In the later case corresponding integrable equation is called $\mathfrak{gl}(2)$ Benjamin-Ono equation. The equation \eqref{ILW-equation} is equivalent to the compatibility condition of the following system (Lax pair)
\begin{subequations}\label{ILW-Lax-pair-lambda}
\begin{align}\label{L-operator-lambda}
   &\bigl((\partial-iv+\lambda)^{2}+u\bigr)\phi^{\scriptscriptstyle{+}}=\lambda^{2}\phi^{\scriptscriptstyle{-}},\\\label{A-operator-lambda}
   &\phi^{\scriptscriptstyle{\pm}}_{t}-i\lambda\phi^{\scriptscriptstyle{\pm}}_{x}-\frac{i}{2}\,\phi^{\scriptscriptstyle{\pm}}_{xx}
   \mp2\phi^{\scriptscriptstyle{\pm}}\,\mathcal{P}_{\scriptscriptstyle{\pm}}v_{x}=0,
\end{align}
\end{subequations}
where $\phi^{\pm}(x)=\phi(x\pm i\tau)$ and $   \mathcal{P}_{\scriptscriptstyle{\pm}}=\frac{1}{2}\left(1\mp i\,\mathcal{T}\right)$.
Lax pair \eqref{ILW-Lax-pair-lambda} admits straightforward generalization for $\mathfrak{gl}(n)$: one has to replace second order differential operator in  \eqref{L-operator-lambda} by $n$-th order.

\bibliographystyle{MyStyle} 
\bibliography{MyBib}
\end{document}